# Spin-filament alignments to unravel galaxy evolution and model intrinsic alignments

Stefania Barsanti[*,1], Clotilde Laigle[2], Nicolas Bouché[3], Anna Rita Gallazzi[4], Mark Sargent[5], Sabine Thater[6], Matthew Colless[7], Scott M. Croom[1], Edward N. Taylor[8], Laurence Tresse[9]

[1]*Sydney Institute for Astronomy (SIfA), School of Physics, The University of Sydney, NSW 2006, Australia*

[2]*Institut d'Astrophysique de Paris, UMR 7095, CNRS, and Sorbonne Université, 98 bis boulevard Arago, 75014 Paris, France*

[3]*Univ Lyon1, Ens de Lyon, CNRS, Centre de Recherche Astrophysique de Lyon (CRAL) UMR5574, F-69230 Saint- Genis-Laval, France*

[4]*INAF-Osservatorio Astrofisico di Arcetri, Largo Enrico Fermi 5, 50126, Firenze, Italy*

[5]*Laboratoire d'Astrophysique (LASTRO), EPFL SB IPHYS, Route de la Sorge, CH-1015, Lausanne, Switzerland*

[6]*Department of Astrophysics, University of Vienna, Turkenschanzstrasse 17, 1180, Vienna, Austria*

[7]*Research School of Astronomy and Astrophysics, Australian National University, Canberra, ACT 2611, Australia*

[8]*Centre for Astrophysics and Supercomputing, Swinburne University of Technology, Melbourne, Australia*

[9]*Aix Marseille Univ, CNRS, CNES, LAM, Marseille, France*

[*]Email: stefania.barsanti@sydney.edu.au

# Abstract

By the 2040s, several all-sky surveys will have transformed our view of the large-scale structure. However, one of the major outstanding questions in astrophysics will remain: understanding how galaxies acquire and evolve their angular momentum and how this connects to the cosmic web. Measuring the alignments between galaxy spins and cosmic filaments across cosmic time, and understanding what this reveals about galaxy evolution, requires surveys that also characterise intrinsic alignments, i.e. correlations in galaxy shapes produced by the cosmic web itself rather than by lensing. Intrinsic alignments are a major source of systematic error in weak-lensing measurements of the fundamental parameters of the Universe. Addressing both questions together will necessitate new types of MOS surveys that combine kinematic information with high-completeness redshifts down to at least 24-25mag.

To achieve our science goals, we require a new generation of wide-field spectroscopic facilities that can obtain spin-filament alignment measurements for millions of galaxies while simultaneously delivering sub-Mpc resolution of the cosmic web and spatially-resolved kinematics required to map the spin–filament connection at the level of individual galaxies within their local cosmic environment. Such a program would provide a unique legacy survey of galaxies and cosmic structures from kiloparsec to megaparsec scales, establishing ESO's leadership in bridging the physics of galaxy evolution with the systematic-control requirements for Stage-IV cosmological surveys.

# Context

In the ΛCDM paradigm, dark matter halos acquire spin through tidal torques during collapse, leading to an alignment between galaxy spin and the orientation of the filament in which the galaxy resides. This alignment is further shaped by the physics of galaxy formation, angular momentum acquisition, and the stages of dynamical evolution tracing galaxy transformation.

Hydrodynamic simulations and observations indicate that the orientation between galaxy spins and filaments is mass dependent (Laigle et al. 2015, Codis et al. 2015). Low-mass, star-forming galaxies tend to align *parallel* to filaments, consistent with growth dominated by smooth gas accretion, whereas more massive, bulge-dominated galaxies show *perpendicular* alignments associated with merger-driven angular momentum reorientation (e.g., Tempel et al. 2013, Codis et al. 2015, Barsanti et al. 2022, Tudorache et al. 2025). Different galaxy formation models predict different flip masses (e.g., Krolewski et al. 2019, Kraljic et al. 2019, Ganeshaiah Veena et al. 2019). This makes galaxy spin–filament alignments a powerful tracer of galaxy assembly history and an observational test of competing galaxy formation models. Cosmological simulations predict a redshift evolution of spin–filament alignments (Codis et al., 2018). Observing this redshift evolution, at least up to $z \sim 1$, is crucial: it provides a direct tracer of how galaxies acquire, lose, and reorient their angular momentum across cosmic time as a function of mass and environment, and therefore constrains the physical conditions that shape galaxy evolution.

# Scientific Rationale

Understanding galaxy spin orientation and amplitude is essential for tracing how galaxies acquire angular momentum, undergo morphological transformations, and interact with the cosmic web (Storck et al. 2025). But beyond galaxy evolution theory, galaxy spin–filament alignments have far-reaching cosmological implications since galaxy spins and shapes both respond to the tidal field of the cosmic web, spin–filament alignments correlate with the shape–filament alignments (Laigle et al. 2025) that drive intrinsic alignments (IAs), a major systematic in weak-lensing (Chisari 2025). IAs mimic lensing signals, distorting estimates of key cosmological parameters such as the total neutrino mass (Lee et al. 2020) and the dark energy equation of state (Lee & Libeskind 2020). Current IA models rely on empirical shape–density correlations and lack physical predictions tied to spin-filament theory.

By combining direct measurements of spin–filament alignments with their associated shape–filament correlations, we can construct physically grounded IA models that predict alignment behaviour across mass, morphology, environment and cosmic time. Such models are essential for addressing beyond-2040s scientific challenges in both galaxy evolution and precision cosmology, where IAs remain a leading systematic in Stage-IV weak-lensing surveys.

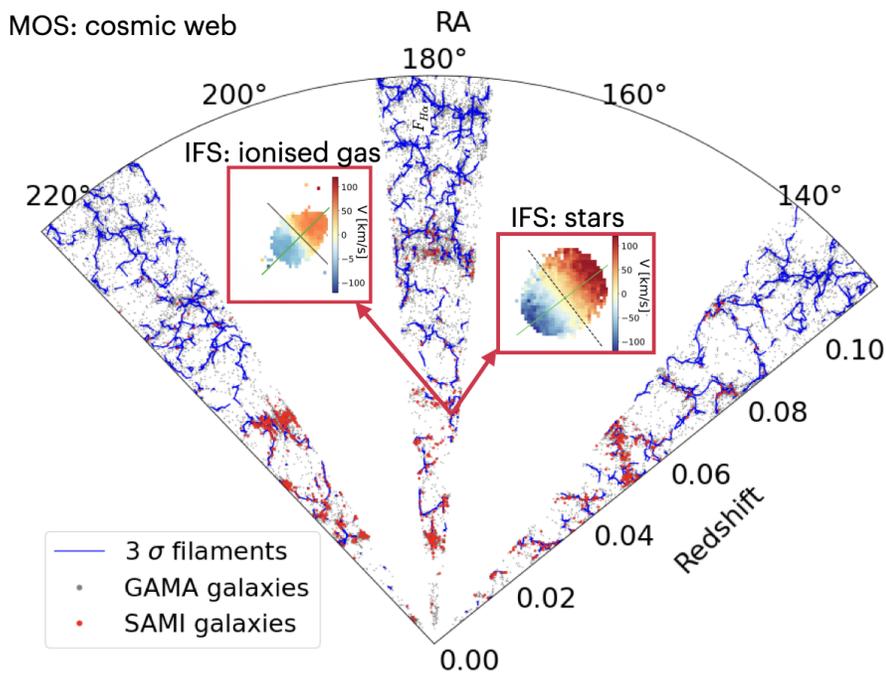

**Figure 1:** Projected network of filaments (blue lines) reconstructed from the GAMA spectroscopic redshift survey (grey points; the most complete dataset today), and 1121 SAMI galaxies with spatially-resolved ionised gas and stellar velocity maps, limited to $z$ = 0.1 (Barsanti et al. 2023). The combination of wide-field MOS for large-area redshift mapping and a monolithic IFS for both spatially-resolved kinematics and high-density spectroscopic sampling tracing the sub-Mpc cosmic web will enable spin–filament alignment studies for millions of galaxies up to $z \sim 1$ while tracing filament spines, measuring local anisotropic tidal fields, and identify small-scale variations in filament thickness and connectivity that drive spin–filament torques.

A facility that combines wide-field MOS for mapping the large-scale environments of millions of galaxies with a monolithic IFS that captures the cosmic web at the full galaxy density across its field *and* resolves ionised-gas *and* stellar kinematics will be able to unlock the physics encoded in these alignments from kpc-to-Mpc scales, enabling the tracing filament spines, local anisotropic tidal fields, variations in filament thickness, and connectivity.

## Key requirements

In order to (i) map the redshift evolution of spin–filament alignments at least up to $z \sim 1$, (ii) to quantify the dependence on mass, morphology, and environment, isolating how feedback, mergers, and filamentary inflows influence spin–filament alignments trends; and (iii) to establish physical empirical models for intrinsic alignments anchored in observed statistics, building predictive models of IAs (rather than empirical corrections), which are crucial for the precision of Stage-IV cosmological parameters, one would need to have a survey with

1. high spectroscopic completeness (>80%) to 24.5 mag and
2. high-resolution (<1 Mpc) cosmic web reconstruction, combined with spatially-resolved ionised gas and stellar kinematics to trace angular-momentum growth.

Given that filaments are 5-20 Mpc in length and 0.1-0.5Mpc in width, these goals require an observing strategy that spans large areas, covering several tens of sq. degrees, AND reach sub-Mpc resolution of the cosmic structures along with spatially-resolved kinematics. The sub-Mpc resolution corresponds to ~40,000 sources/sq.degree with redshifts (10-12 per sq. arcmin). The resulting dataset will be able to deliver the most comprehensive map of the cosmic web, revealing the baryon cycling in galaxies and providing physical IA priors to reduce systematic uncertainties in cosmological parameters, and become a defining legacy archive for the community, enabling cross-disciplinary studies of extragalactic astrophysics and cosmology.

## Limitations of 2020s/2030s facilities

With current facilities, the measurements of the redshift dependence of spin–filament alignments will remain out of reach for at least the next 15 years due to fundamental limitations in survey scale, target densities, kinematic coverage, and redshift reach of the current generation observatories. Planned observations lack the necessary combination of large sample size (≳$10^6$ galaxies), spatially-resolved kinematics, and the redshift range required to test theoretical predictions for their redshift evolution and mass-dependent transitions. Such a large sample size is essential to comprehensively trace galaxy transformation in relation to their spin alignment within the cosmic web. Splitting galaxies by stellar mass, morphology, large-scale overdensity, and distance to filaments, nodes, and voids requires statistically significant numbers in each bin. Only with these large, well-populated subsamples can we physically disentangle the mechanisms driving spin–filament alignments.

Current observational efforts are limited given that imaging surveys (e.g., SDSS, KiDS, COSMOS, Euclid) offer large galaxy samples (~$10^9$) with high projected density in tomographic slices but lack spectroscopic redshifts and spin information since they can only characterize photometric shapes. In contrast, IFS data (e.g., MaNGA, SAMI, Hector, MAGPI) provide spin measurements, but only for small samples (~$10^3$-$10^4$), narrow sky coverage ($\lesssim$ few hundred deg²), low target densities ($\lesssim$10–50 galaxies deg$^{-2}$), and low redshifts ($z \lesssim 0.3$).

Euclid, Rubin, Roman, SKA and the ELT will transform our view of galaxy evolution and large-scale structure, but none of them will deliver the required combination of wide-area multiplexed redshift mapping, resolved galaxy kinematics, and sub-Mpc reconstruction of filaments at the full galaxy density, which together are needed to trace how galaxies acquire and reorient their angular momentum within the evolving cosmic web across cosmic time.

Large-multiplex MOS surveys such as DESI, 4MOST, PFS, and MOONS will deliver unprecedented spectroscopic statistical samples but reach effective target densities of only a few galaxies per square arcminute, set by hard limits in fibre positioning and multiplex design. Yet, the galaxy surface density up to $z \sim 1$ is 30–50 times higher. As a result, MOS surveys undersample the cosmic web, preventing the recovery of filament spines, thickness, and local tidal anisotropy at the sub-Mpc resolution required to detect spin–filament alignment signatures. Existing or planned IFU instruments, such as ELT/HARMONI, ELT/MOSAIC, VLT/MUSE, AAT/SAMI, AAT/Hector, Keck/KCWI, and JWST/NIRSpecIFU, cover areas far too small to map the cosmic web or build statistically representative galaxy samples spanning mass, environment, and redshift.

This leaves a fundamental gap: no current or planned facility can simultaneously deliver (1) millions of galaxy MOS redshifts, (2) high enough galaxy sampling density for sub-Mpc filament reconstruction out to $z \sim 1$, and (3) well-resolved kinematic maps necessary to anchor physically predictive models of angular momentum and IAs. Consequently, the key physical link between gas accretion, stellar assembly, and large-scale torques will remain poorly constrained for another generation.

## Survey requirements for a new facility

To achieve the sensitivity and spatial resolution needed for kinematic measurements out to $z \sim 1$, where the key ionised-gas tracers of angular momentum (e.g. [O II] 3727) become faint, a 12-m ground-based facility is required. A wide-field MOS with a field of view of at least 3 deg² is essential to capture representative volumes of the cosmic web, sampling filaments, nodes and voids simultaneously, and to achieve the sky coverage required for statistically significant measurements across stellar mass, morphology, and environment. A high-multiplex spectrograph is needed to obtain redshifts. A monolithic IFS with a 3x3 arcmin² field of view would map spin–filament alignments across cosmic time at an unprecedented resolution, anchor physically predictive IA models, and deliver the statistical and kinematic precision essential for Stage-IV cosmology. The proposed Wide-field Spectroscopic Telescope (WST; Mainieri et al. 2024) facility, which combines all these capabilities simultaneously, would be well-suited for this survey and to achieve these science goals.